\documentclass[epj,referee]{svjour}
\usepackage{graphicx}
\usepackage{amssymb,amsmath,amssymb}

\begin{document}

\title{Modulated class A laser: Stochastic resonance in a limit-cycle potential system}

\author{Catalina Mayol, Ra\'{u}l Toral and Horacio S. Wio}
\institute{IFISC (Instituto de F{\'\i}sica Interdisciplinar y Sistemas Complejos), Universitat de les Illes Balears-CSIC, 07122-Palma de Mallorca, Spain}
\date{\today}
\authorrunning{C. Mayol, R. Toral, H.S. Wio}
\titlerunning{Modulated class A laser}
\abstract{
We exploit the knowledge of the nonequilibrium potential in a model
for the modulated class A laser. We analyse both, the deterministic
and the stochastic dynamics of such a system in terms of the
Lyapunov potential.
Furthermore, we analyse the stochastic response of such a system and
explain it again using the potential in a wide range of parameters
and for small values of the noise. Such a response is quantified by
means of the amplification factor, founding stochastic resonance
within specific parameter's ranges.
}
\maketitle

\section{Introduction}

Stochastic resonance (SR) has become a paradigm of the constructive effects of fluctuations on
nonlinear systems~\cite{sr-01,sr-02}. Briefly, the phenomenon occurs whenever the Kramers' rate
for the transition between attractors matches the typical frequency of a signal which can not
itself trigger that transition (i.e. it is sub-threshold. However, supra-threshold cases have also
been studied~\cite{abott}). Several measures of SR can be
defined (the signal-to-noise ratio and the spectral amplification factor being the
most used ones), and the theoretical analysis is usually carried on in terms of the two-state
approximation~\cite{sr-01}. Since its discovery more than thirty years ago interest has gradually shifted towards
increasingly complex systems, networks and nonlinear media being the main directions. Instances
of this trend are the experiments carried out to explore the role of SR in sensory and other
biological functions~\cite{sr-04}, and experiments in chemical systems~\cite{sr-05}.

Recent research was based on the study of nonlinear media that can be described
as reaction-diffusion systems, namely those that can be thought of as a collection
of diffusively coupled nonlinear units. The possibility of enhancing the system's
response through the coupling of those units~\cite{sr-06,sr-07,sr-08,sr-09,sr-10,sr-11,sr-12}
has been among the explored issues, together with the problem of how does nature manage to make
the system's response less dependent on a fine tuning of the noise intensity, or that of searching
for different ways to control the phenomenon~\cite{sr-13,sr-14}.

In the case of extended systems, the theoretical studies have profited from
the knowledge of the system's \textit{nonequilibrium potential} (NEP)
~\cite{Graham,sr-15}. However, there are models that realistically describe
experimental systems, with a known form of the NEP, that have so far not been
exploited for SR's studies. One such a system is the \textit{modulated class A
laser with injection}~\cite{laser0,laser1}.

The dynamics of such lasers have a simple description
in terms of rate equations for the temporal evolution
of the different dynamical variables. It is usual to classify
lasers according to the decay rate of photons, carriers, and
material polarisation~\cite{las1,las2,las3}. In the so-called
class A lasers
the material variables decay to the steady state much faster
than the electric field, and can therefore be adiabatically
eliminated. The resulting equation for the electric field suffices
to describe the dynamical evolution of the laser. This
equation contains a white-noise term accounting for the stochastic
nature of the spontaneous emission. Some properties
of typical class A lasers, such as a dye laser, are discussed in
Refs.~\cite{las4,las5}.

Here we analyse SR in the indicated laser system, exploiting the knowledge of
its NEP. The extremely interesting
aspect of this system's NEP is that it goes from a fixed point into a limit
cycle potential as a parameter of the laser is varied. In the following Section 2 we
introduce the model and the form of its NEP. In Section 3 we describe the dynamics of
the system when modulation is considered, both in the deterministic and the stochastic
situations. Afterwards, we devote section 4 to the study of the amplification factor
to have a deep understanding of the stochastic dynamics with modulation, therefore
we discuss a few situations of interest. Finally, we draw some conclusions.

\section{The model}
We study the dynamical equations of a class A laser described in terms of the slowly varying complex amplitude $E$ of the electric field, injected with a monochromatic optical field $S e^{i \Omega t}$. The resulting evolution equation is~\cite{laser1,haken,vandergraaf}
\begin{equation}
\dot{E}(t)=(1+i\alpha)\left(\frac{\Gamma}{1+ \beta |E|^2}- \kappa \right) E + \sigma S e^{-i \Delta\Omega t} + \zeta(t), \label{cAm:camp}
\end{equation}
where $\Delta \Omega$ is the detuning between the external field and the free-running laser frequency. Here $\kappa$ is the cavity decay rate, $\Gamma$ the gain parameter, $\beta$ the saturation-intensity parameter, $\alpha$ the atomic detuning parameter, and $\sigma$ the amplitude feed-in rate, proportional to the inverse of the round-trip time $\tau_\textrm{in}$~\cite{Agrawal}, $\zeta(t)$ is a (complex) Gaussian white-noise term with zero mean and correlations $\langle\zeta(t)\zeta^*(t')\rangle=4D\delta(t-t')$. In this paper we will work with a dimensionless version of Eq.(\ref{cAm:camp}) with rescaled time and electric field. Writing $x_1+ix_2$ for the new rescaled field, the new equations are
\begin{eqnarray}
\dot{x}_1 & = & \left(\frac{a}{b+x_1^2+x_2^2} -1 \right)(x_1- \alpha x_2) + \rho - \eta x_2 + \sqrt{2\epsilon}\xi_1(t), \nonumber\\ \label{cAm:x1}
\\
\dot{x}_2 & = & \left(\frac{a}{b+x_1^2+x_2^2} -1 \right) (\alpha x_1 + x_2) + \eta x_1 + \sqrt{2\epsilon}\xi_2(t), \label{cAm:x2}
\end{eqnarray}
(see Ref.~\cite{laser1} for full details). The new parameters are $a=\Gamma/(\kappa \beta)$ (related to the gain parameter), $b=1/\beta$, $\rho=\sigma S / \kappa$ and $\eta=\Delta\Omega/\kappa$ such that $\rho$ is proportional to the intensity of the injected field and $\eta$ to its frequency. The (real) Gaussian white noises $\xi_i(t)$ have zero mean and correlations $\langle\xi_i(t)\xi_j(t')\rangle=\delta(t-t')\delta_{ij}$, $i,j=1,2$,
with noise intensity $\epsilon=D/\kappa$. The equations can be written in terms of intensity $I=x_1^2+x_2^2$ and phase $\phi=\arctan(x_2/x_1)$ as
\begin{eqnarray}
\dot I&=&2\left[\frac{a}{b+I}-1\right]I+2\rho\sqrt{I}\cos(\phi)+2\sqrt{2\epsilon I}\xi_I(t),\label{eqdotI}\\
\dot \phi&=&\alpha\left[\frac{a}{b+I}-1\right]-\frac{\rho}{\sqrt{I}}\sin(\phi)+\eta+\frac{\sqrt{2\epsilon}}{\sqrt{I}}\xi_\phi(t),\label{eqdotphi}
\end{eqnarray}
with white noises of zero mean and correlations\\ $\langle\xi_A(t)\xi_B(t')\rangle=\delta(t-t')\delta_{AB}$, $A,B=I,\phi$.

Eventually, in order to study the existence of stochastic resonance in this system, we will be interested in allowing the parameter $a$
to vary periodically in time. However, when $a$ is a constant
it is a matter of simple algebra to show that the dynamical Eqs.(\ref{cAm:x1},\ref{cAm:x2}) can be written as~\cite{sr-15,laser1,hsw,wdl,Gardi}
\begin{equation}
\begin{pmatrix}\dot x_1\\ \dot x_2\end{pmatrix}=\displaystyle-{\cal M}\begin{pmatrix}\frac{\partial V}{\partial x_1}\\ \frac{\partial V}{\partial x_2}\end{pmatrix}+\eta\begin{pmatrix} -x_2 \\x_1 \end{pmatrix}+\begin{pmatrix} \xi_1 \\ \xi_2 \end{pmatrix},\label{matrix}
\end{equation}
with the matrix ${\cal M}=\begin{pmatrix}1&-\alpha\\ \alpha&1\end{pmatrix}$, and the potential function
\begin{eqnarray}
V(x_1,x_2)&=&\frac{1}{2} \left[x_1^2+x_2^2-a \ln(b+x_1^2+x_2^2)\right] \nonumber\\
&& -\frac{\rho}{(1+\alpha^2)}(x_1-\alpha x_2) \label{vx}.
\end{eqnarray}
For future reference, we write this function in terms of intensity $I$ and phase $\phi$,
\begin{equation}
V(I,\phi)= \frac{1}{2}\left[I-a \ln(b+I)\right] - \frac{\rho \sqrt{I}}{\sqrt{1+\alpha^2}}\cos(\phi+\arctan(\alpha)).
\label{vi}
\end{equation}
In the next subsections we summarise the main features of the dynamical evolution in the deterministic and stochastic cases, a more detailed account can be found in~\cite{laser1}.

\subsection{Deterministic dynamics}
In the absence of the stochastic terms ($\epsilon=0$), it turns out that $V(x_1,x_2)$ is a Lyapunov potential (i.e. bounded from below and never increasing during the dynamical evolution) provided that the condition $\eta\rho=0$ is satisfied. The existence of the potential allows one to ``visualise'' the dynamics of the point of coordinates $(x_1(t),x_2(t))$ as the movement of a fictitious particle within the potential landscape.

(i) The case $\rho=0$: If $a<b$ the potential has a single minimum at $x_1=x_2=0$ and the dynamics leads to this only stable fixed point. If $a>b$, the Lyapunov potential has the shape of a ``Mexican hat'' (see for instance Fig.7 in~\cite{laser0})
with a line of minima at the circle $I=x_1^2+x_2^2=a-b$. After a transient time to reach the minima of the potential, the intensity remains constant at $I=b-a$ but there is a residual dynamics $\dot\phi=\eta$ on the potential minimum induced by the terms proportional to $\eta$ which do not vary the value of $V$. As a result, in the stationary state the phase increases linearly as $\phi(t)=\phi_0+\eta t$, with $\phi_0$ depending on the initial conditions.

(ii) The case $\rho\ne 0,\eta=0$: Independently of the values of $a$ and $b$ the potential is now tilted to a particular direction and displays a single minimum. Hence, all trajectories end up on a steady state with a well defined intensity and phase obtained from setting $\dot I=\dot \phi=0$ in Eqs.(\ref{eqdotI},\ref{eqdotphi}).

(iii) In the case $\rho\ne 0,\eta \ne0$, the function $V(x_1,x_2)$ is no longer a Lyapunov potential. However, the complex bifurcation set that appears in this case~\cite{laser1}, can still be understood in terms of the potential at least for small values of the product $\rho\eta$.

In all cases, the details of the transient trajectory in the $(x_1,x_2)$ space leading to the minimum of the potential depend on the particular value of the parameter $\alpha$.

\subsection{Stochastic dynamics}
In the presence of noise terms ($\epsilon \ne 0$), the Lyapunov potential also allows us to determine the steady-state probability distribution as $P_\textrm{st}(x_1,x_2)=Z^{-1}\exp(-V(x_1,x_2)/\epsilon)$, being $Z$ the normalisation constant. The relation is exact if $\eta\rho=0$ and, otherwise, it has to be interpreted as an approximation valid in the limit $\epsilon\to 0$~\cite{Graham}. A simple change of variables gives us the probability distribution for intensity and phase, $P_\textrm{st}(I,\phi)={\hat Z}^{-1}\exp{(-V(I,\phi))/\epsilon}$, from which the average value of intensity $I$ and phase flux, $\dot \phi$, can be computed
\begin{eqnarray}
\label{Ist}
\langle I \rangle_\textrm{st}&=&\int I P_\textrm{st}(I,\phi) dId\phi,\\
\label{phist}
\langle \dot{\phi} \rangle_\textrm{st}&=&\int \left[ \alpha \left(\frac{a}{b+I}-1\right) -\frac{\rho}{\sqrt{I}}\sin(\phi)+\eta \right] P_\textrm{st}(I,\phi) d\phi dI.\nonumber\\
\end{eqnarray}
For $\rho=0$ and $a>b$, the integrals can be analytically computed with the results
\begin{eqnarray}
\label{Ist0} \langle I \rangle_\textrm{st} &=& a-b+ 2 \, \epsilon \,
\left[1 +\frac{\exp(-b/2 \, \epsilon) \, (b/2 \,
\epsilon)^{\frac{a}{2 \, \epsilon}+1}} {\Gamma\left(\frac{a}{2 \,
\epsilon} +1, \frac{b}{2 \, \epsilon}\right)}\right],\\
\label{phist0}
\langle \dot{\phi} \rangle_\textrm{st}&=&- \alpha
\frac{\exp(-b/2 \, \epsilon) \, (b/2 \,
\epsilon)^{\frac{a}{2 \, \epsilon}}} {\Gamma\left(\frac{a}{2 \,
\epsilon}+1, \frac{b}{2 \, \epsilon}\right)} +\eta,
\end{eqnarray}
where $\Gamma(x,y)$ is the incomplete gamma function~\cite{tabla}. From these expressions we derive that, in the deterministic case, $\epsilon=0$ (and $\rho=0\,,a>b$), the average value of the intensity is $\langle I \rangle_\textrm{st} = a-b$ and the phase changes linearly as $\phi(t)=\phi_0+\eta t$, in accordance with the discussion of the previous section. Note that in the case $\rho=0$ the parameter $\alpha$ has no influence in the mean value of the intensity but only in the phase flux. In the most general case $\rho \ne 0$, the steady state average values have to be computed from a numerical integration of Eqs.(\ref{Ist},\ref{phist}).

\section{Modulation: Deterministic and Stochastic Dynamics}

We are interested in analysing the possibility of stochastic resonance in the dynamic system's response to an external perturbation. To this end, we consider a periodic modulation of the parameter $a$ in
Eqs.(\ref{cAm:x1},\ref{cAm:x2}) of the form:
\begin{equation}
a(t)=a_0+a_1 \sin(\omega t).
\label{amod}
\end{equation}
As $a = \Gamma /(\kappa \beta)$, this is equivalent to a
modulation of the gain parameter (with fixed values for the
cavity decay rate and the saturation-intensity parameter).
We firstly consider this contribution in absence
of noisy terms (deterministic) and then in presence of noise (stochastic effects).

\subsection{Modulation: Deterministic Dynamics}

We first note that the form of the evolution Eqs.(\ref{matrix}) is still valid if the parameter $a$ depends on time $a(t)$.

\begin{figure}
\centerline{
\includegraphics[scale=0.5]{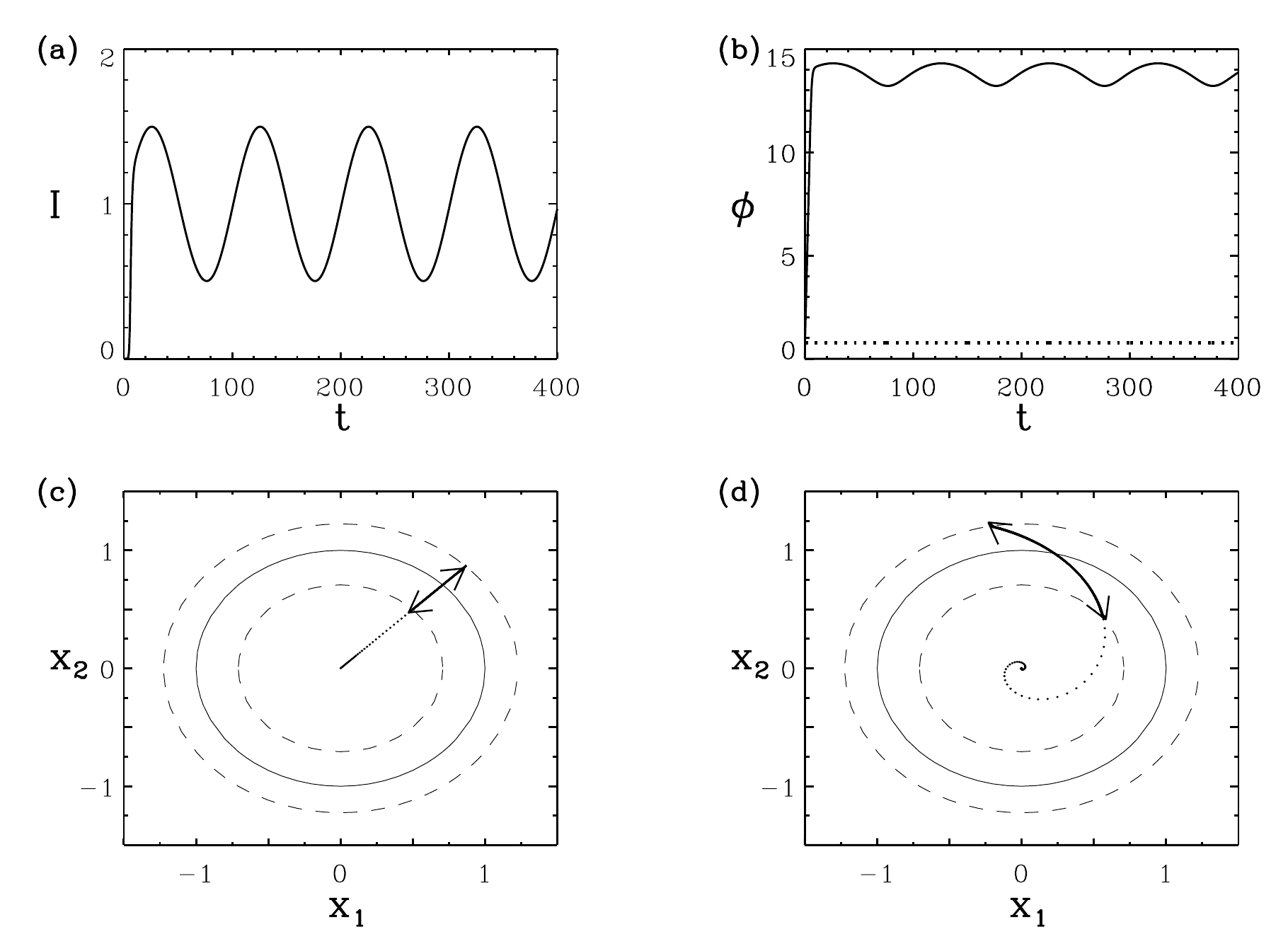}} \caption{Evolution of the intensity $I={x_1}^2+{x_2}^2$, phase $\phi=\arctan(\frac{x_2}{x_1})$ and trajectories in the phase space $(x_1,x_2)$ coming from a numerical integration of Eqs.(\ref{cAm:x1},\ref{cAm:x2}) in the case of no injected signal: $\rho=0$, $\eta=0$ and no noisy terms: $\epsilon=0$. The intrinsic laser parameters are $a_0=2$, $b=1$, while the parameters of the modulation Eq.(\ref{amod}) are: $a_1=0.5$ and $\omega=2\pi /T$, $T = 100$.
(a) Intensity versus time for $\alpha=0$ and $\alpha=2$ (both lines coincide).
(b) Phase versus time for $\alpha=0$ (dotted line) and $\alpha=2$ (solid line).
Dynamics in the ($x_1$,$x_2$) space: (c) $\alpha=0$ and (d) $\alpha=2$. The arrows of panels (c) and (d)
indicate the oscillatory dynamics of the system in the stationary state.
In those same panels the circles ${x_1}^2+{x_2}^2=a_0-b$ (solid circle),
${x_1}^2+{x_2}^2=a_0-b+a_1$ and ${x_1}^2+{x_2}^2=a_0-b-a_1$ (dotted circles) are displayed
in the graphs $x_2$ versus $x_1$. These circles correspond to the extremes and
the middle of the minima of the potential. }\label{laser1}
\end{figure}

\begin{figure}
\centerline{
\includegraphics[scale=0.5]{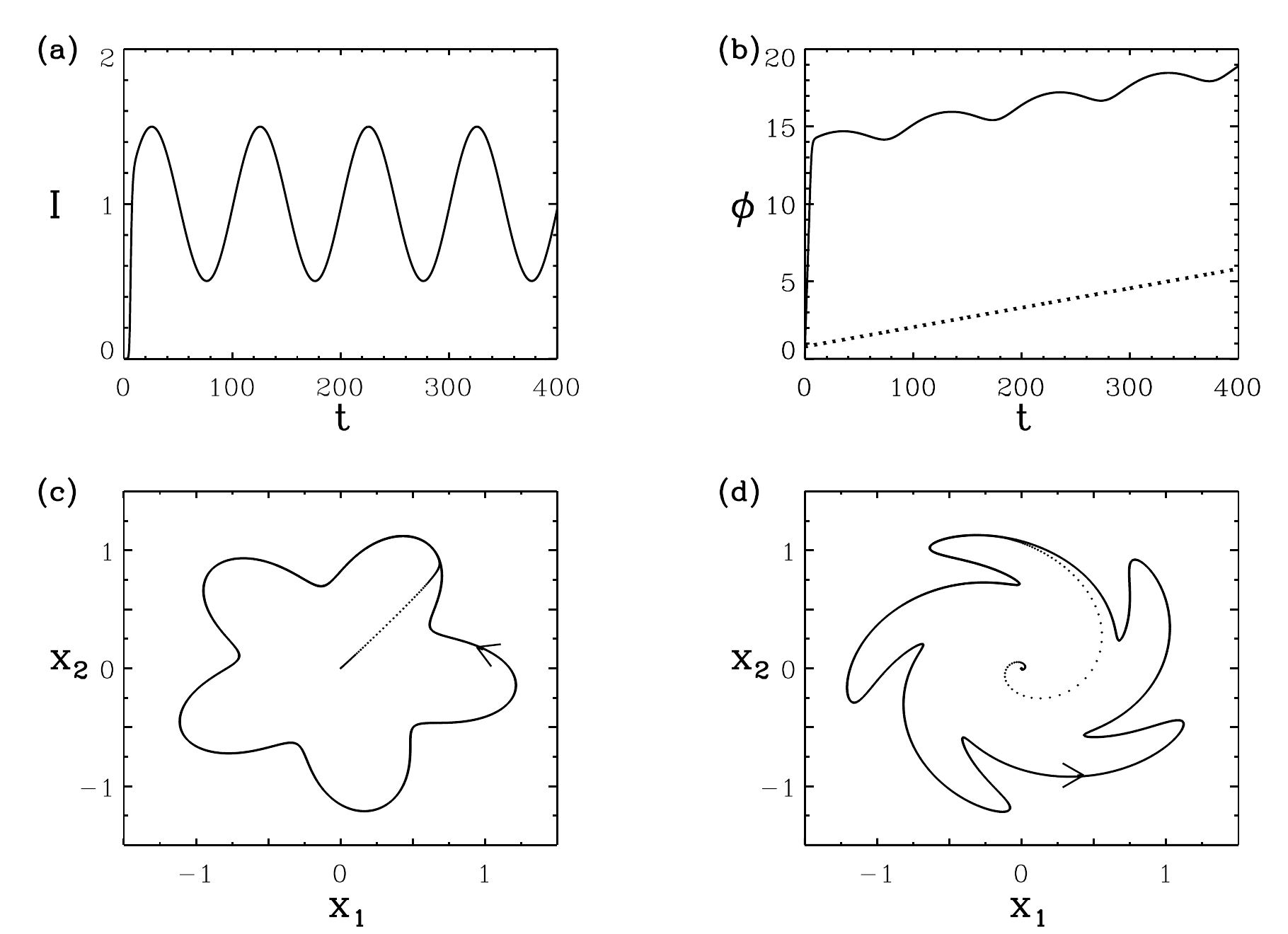}} \caption{Same as Fig.\ref{laser1}, with $\rho=0$, but $\eta=\omega/5$. The main difference, as compared to the case $\eta=0$ shown in Fig.\ref{laser1}, is that the laser phase $\phi(t)$ has now an additional dependence $\eta t$ and the trajectories oscillate around the centre of coordinates in the $(x_1,x_2)$ plane. As the ration $\omega/\eta$ is an integer number the trajectories are closed in that plane both for $\alpha=0$, panel (c), and $\alpha=2$, panel (d).}\label{laser2}
\end{figure}

\begin{figure}
\centerline{
\includegraphics[scale=0.5]{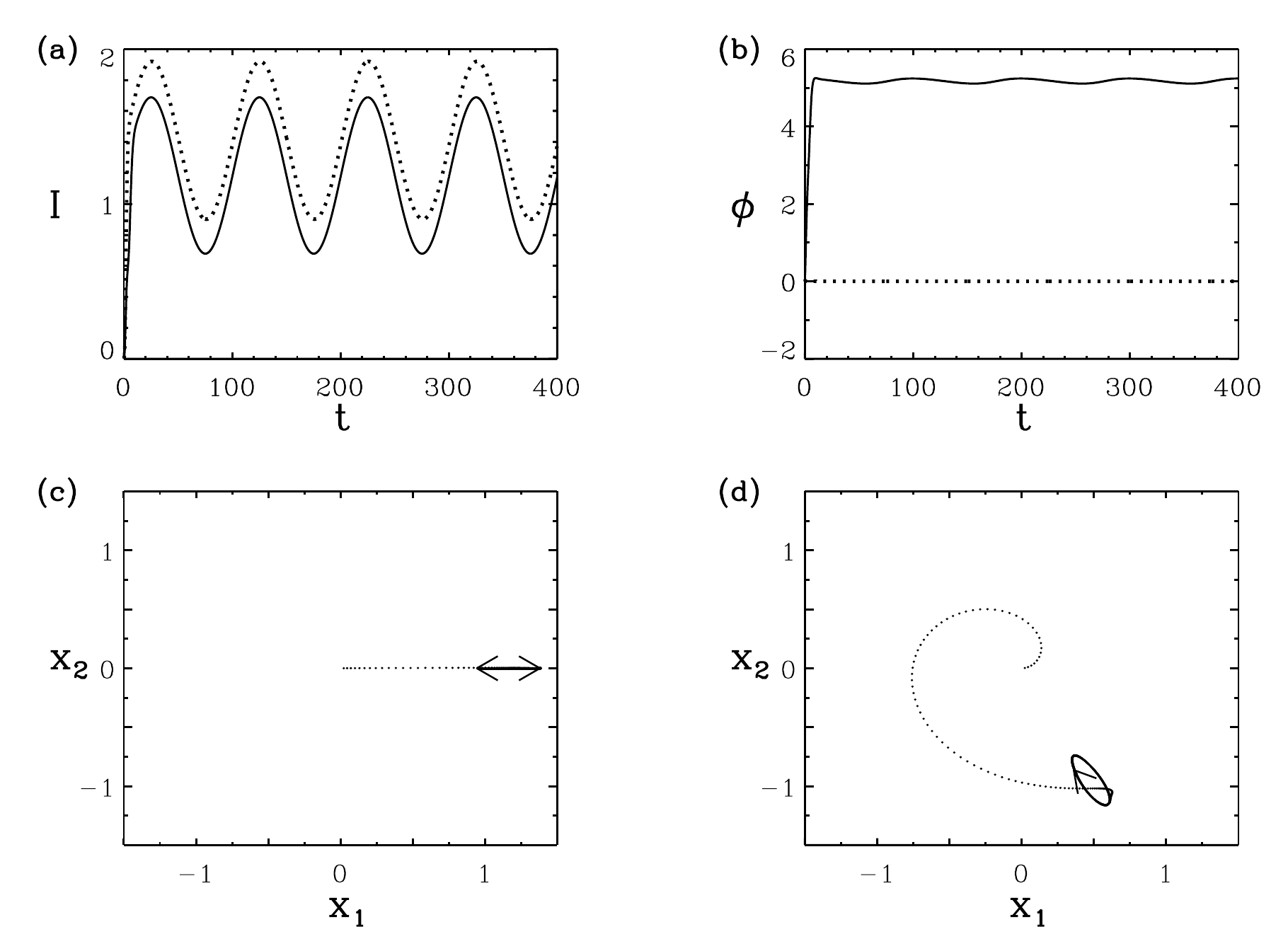}} \caption{Same as in Fig.\ref{laser1} with an injected signal: $\rho=0.2$, $\eta=0$. At variance with Fig.\ref{laser1} the $\alpha$ parameter has now effects on the oscillations of the intensity $I(t)$: $\alpha=0$ (dotted line) and $\alpha=2$ (solid line). For $\alpha=0$, it is $\phi=0$ in the stationary regime, dotted line of panel (b) and the trajectory oscillates around the $x_2=0$ line, see arrowed segment of panel (c), while they follow a more complicated closed trajectory in the case of $\alpha=2$, panel (d).}\label{laser3}
\end{figure}

For the deterministic dynamics, i.e. $\epsilon=0$, the numerical results indicate that in the stationary state (long times), the intensity $I(t)$ varies with time, with the same frequency of the modulation term, $\omega$, see Figs.\ref{laser1},\ref{laser2},\ref{laser3}. This modulated behaviour can still be understood with the use of the potential in Eq.(\ref{vi}) replacing $a$ by a time-dependent modulation as given by Eq.(\ref{amod}). Note that this potential has now an explicit dependence on time, i.e $V(I,\phi;t)$.

We keep in this section the condition $a_0>b$ such that the unmodulated ($a_1=0$) potential $V(x_1,x_2)$ has the ``Mexican hat'' shape with a line of degenerate minima at $x_1^2+x_2^2=a_0-b$.

(i) For $\rho=0$ and $a_0>a_1+b$ the modulated potential keeps at all times the same qualitative shape with a time-dependent minimum $I_\textrm{min}(t)= a_0-b +a_1 \sin(\omega t)$. In this case, it is observed that for large modulation periods, small values of $\omega$, and after a transient time, the trajectories $(x_1(t),x_2(t))$ follow faithfully that minimum but with a time delay, such that their intensity $I(t)=x_1(t)^2+x_2(t)^2$ after this transient time can be fitted as $I_\textrm{st}(t)=a_0-b + a_1 \sin(\omega t + \psi)$, with $\psi$ a constant angle, see panel (a) of Fig.\ref{laser1} corresponding to $\eta=0$. However, for small modulation period --large frequency $\omega$-- the shape of the potential changes very fast and, although the trajectories $(x_1(t),x_2(t))$ tend to the minima of the potential following the maximum slope lines, they are not able to follow adiabatically the values of these minima. As a result, it turns out that the intensity can be fitted to a form $I_\textrm{st}(t)=a_0-b+c_1\sin(\omega t + \psi)$, with $c_1< a_1$ (not shown).

For $\alpha=0$, the phase $\phi(t)$ remains constant around a value that depends on the initial condition, see dotted line of Fig.\ref{laser1}(b), also manifested by a oscillatory trajectory in space $(x_1, x_2)$, as indicated by the arrowed straight segment in Fig.\ref{laser1}{(c)}. A non zero value of $\alpha$ induces a periodic variation of the phase $\phi(t)$ with the same frequency $\omega$ of the external modulation, see solid line in Fig.\ref{laser1}{(b)} and a modification of the trajectories in the space $(x_1, x_2)$ that, although still oscillatory, do not fall on a straight segment as indicated by the arrowed curved line in Fig.\ref{laser1}(d).

The main effect of a non-zero value of the external injection frequency $\eta$ (while still keeping $\rho=0$) is to increase the phase in an amount $\eta t$ with respect to the value for $\eta=0$, while keeping the same evolution for the intensity $I(t)$. Compare Fig.\ref{laser1} for $\eta=0$ and Fig.\ref{laser2} for $\eta=\omega/5$. On the $(x_1,x_2)$ plane, the trajectories now oscillate around the centre of coordinates, the exact shape depending on the value of $\eta$: if (as displayed in the figure) $\omega/\eta$ is an integer number, the trajectories form a closed loop.

(ii) For $\rho > 0,\,\eta=0$ the potential is tilted to a preferred direction and displays a single minimum whose location oscillates periodically in time with the frequency $\omega$. The trajectories in the $(x_1,x_2)$ plane as well as the intensity and phase follow this time-varying minimum as shown in Fig.\ref{laser3}. Note that the stationary intensity oscillates around a larger mean value than $a_0-b$, Fig.\ref{laser3}(a). The maximum and minimum value of the oscillations of the intensity $I(t)$ now depend on the $\alpha$ parameter at variance with the case $\rho=\eta=0$ analysed previously.

The combined effects of $\eta$ and $\rho$ include different scenarios depending on the values adopted for these two parameters, corresponding to an extension of the bifurcation set reported in the non-modulated case.

\subsection{Modulation: Stochastic Effects}

We now consider Eqs.(\ref{cAm:x1},\ref{cAm:x2}) subject simultaneously to the modulation term (\ref{amod}) and to noise (i.e. $\epsilon \ne 0$). The noise term $\epsilon$ is the responsible of driving the system out of the minima of the potential. In this section we will discuss the possible enhancement of the modulation induced by the stochastic terms as a function of the different system parameters. It is our main aim here to characterise the stochastic dynamics in terms of the Lyapunov potential. We use the asymptotic probability distribution function $P_\textrm{st}(I,\phi,t)= {\hat Z}^{-1}\exp{(-V(I,\phi,t))/\epsilon})$ where the potential $V(I,\phi,t)$ is given by Eq.(\ref{vi}) but including an explicit dependence of $a(t)$ with time as given by Eq.(\ref{amod}).

\begin{figure}
\centerline{
\includegraphics[scale=0.5]{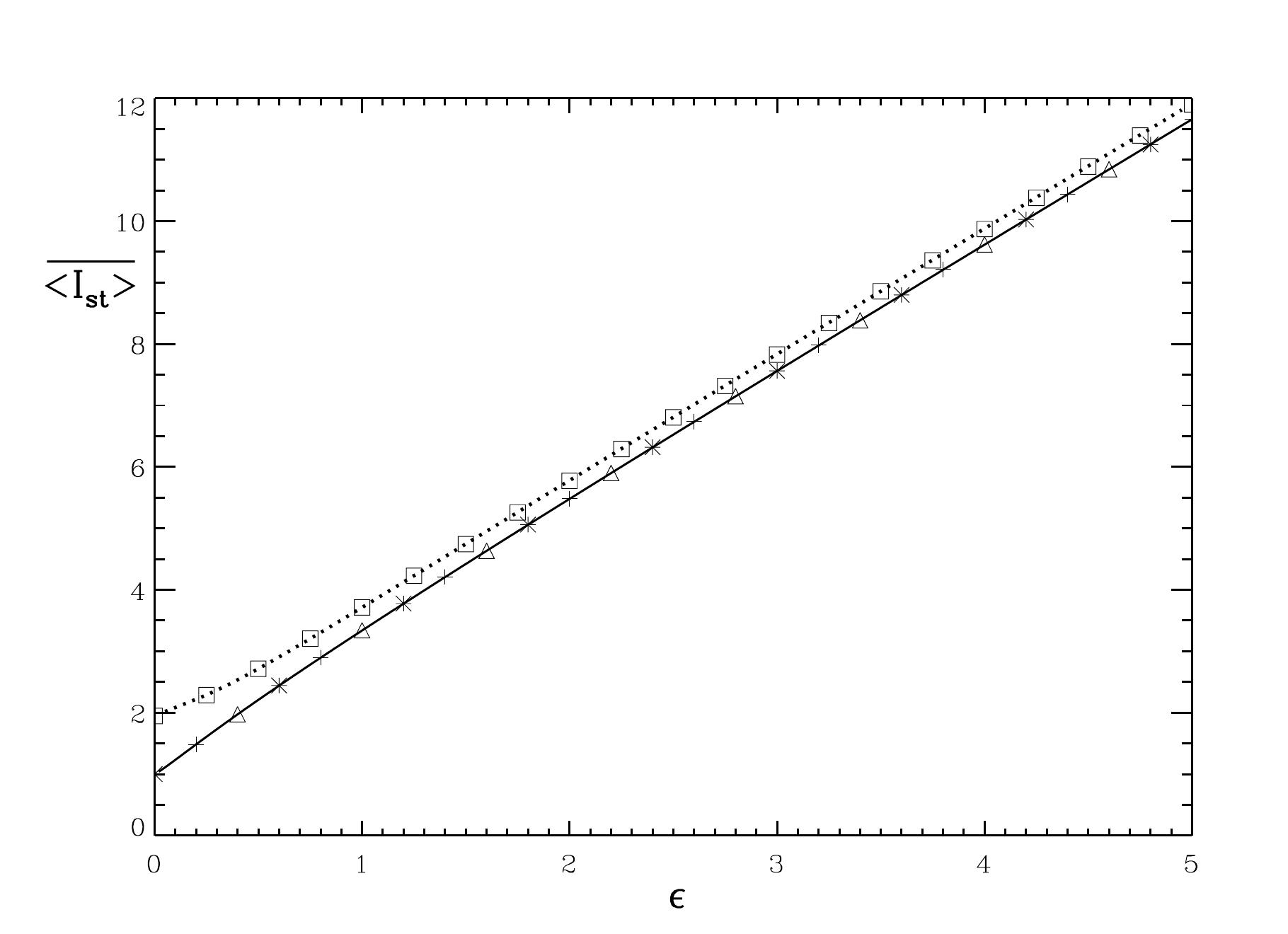}} \caption{Plot of $\overline{\langle I \rangle_\textrm{st}}$
 as a function of the noise intensity $\epsilon$. Common parameters $a_0$, $b$, $a_1$ and $\omega$ as in previous Fig.\ref{laser1}. Lines correspond to expression (\ref{imean}): solid line for $\rho=0$ (results depend neither on $\alpha$ nor on $\eta$), dotted line for $\rho=1$. The symbols are the results of averaging over at time $10^5 T$ the data for $I(t)$ coming from the numerical integration of Eqs.(\ref{cAm:x1},\ref{cAm:x2}). Symbols: $(+)$ $\alpha=-5$, $\eta=0$ and $\rho=0$,
 $(\ast)$ $\alpha=2$, $\eta=0$ and $\rho=0$, $(\triangle)$ $\alpha=2$, $\eta=0.5$ and $\rho=0$, $(\square)$ $\alpha=2$, $\eta=0$ and $\rho=1$.}
\label{figureintensity}
\end{figure}

We can obtain the mean values of the intensity and phase flux in a similar way than in Eqs.(\ref{Ist}, \ref{phist}). As these values have an explicit dependence on time, we obtain the mean value of each of them in a period of time, i.e.
\begin{equation}
\overline{\langle I \rangle_\textrm{st}}= \frac{\omega}{2\pi} \int_0^{\frac{2\pi}{\omega}} \langle I \rangle_\textrm{st}(t) dt.
\label{imean}
\end{equation}
and
\begin{equation}
\overline{\langle \dot{\phi} \rangle_\textrm{st}}= \frac{\omega}{2\pi} \int_0^{\frac{2\pi}{\omega}} \langle \dot{\phi} \rangle_\textrm{st}(t) dt.
\label{dphasemean}
\end{equation}
where the averages $\langle I \rangle_\textrm{st}(t)$ and $\langle \dot{\phi} \rangle_\textrm{st}(t)$ are obtained from Eqs.(\ref{Ist},\ref{phist}) using the time dependent probability distribution $P_\textrm{st}(I,\phi,t)$. We note that Eqs.(\ref{imean}, \ref{dphasemean}) do not have a dependence on $\omega$, i.e. by defining a new time $t'=\omega t$ the variable $\omega$ disappears from both integrals. Furthermore, $\overline{\langle I \rangle_\textrm{st}}$ neither depends on $\alpha$ nor on $\eta$. A result that is confirmed in Fig.\ref{figureintensity} where the lines coming from the theoretical expressions (solid line for $\rho=0$ and dotted line for $\rho=1$) do not change for different values of $\alpha$ or $\eta$.

\begin{figure}
\centerline{
\includegraphics[scale=0.5]{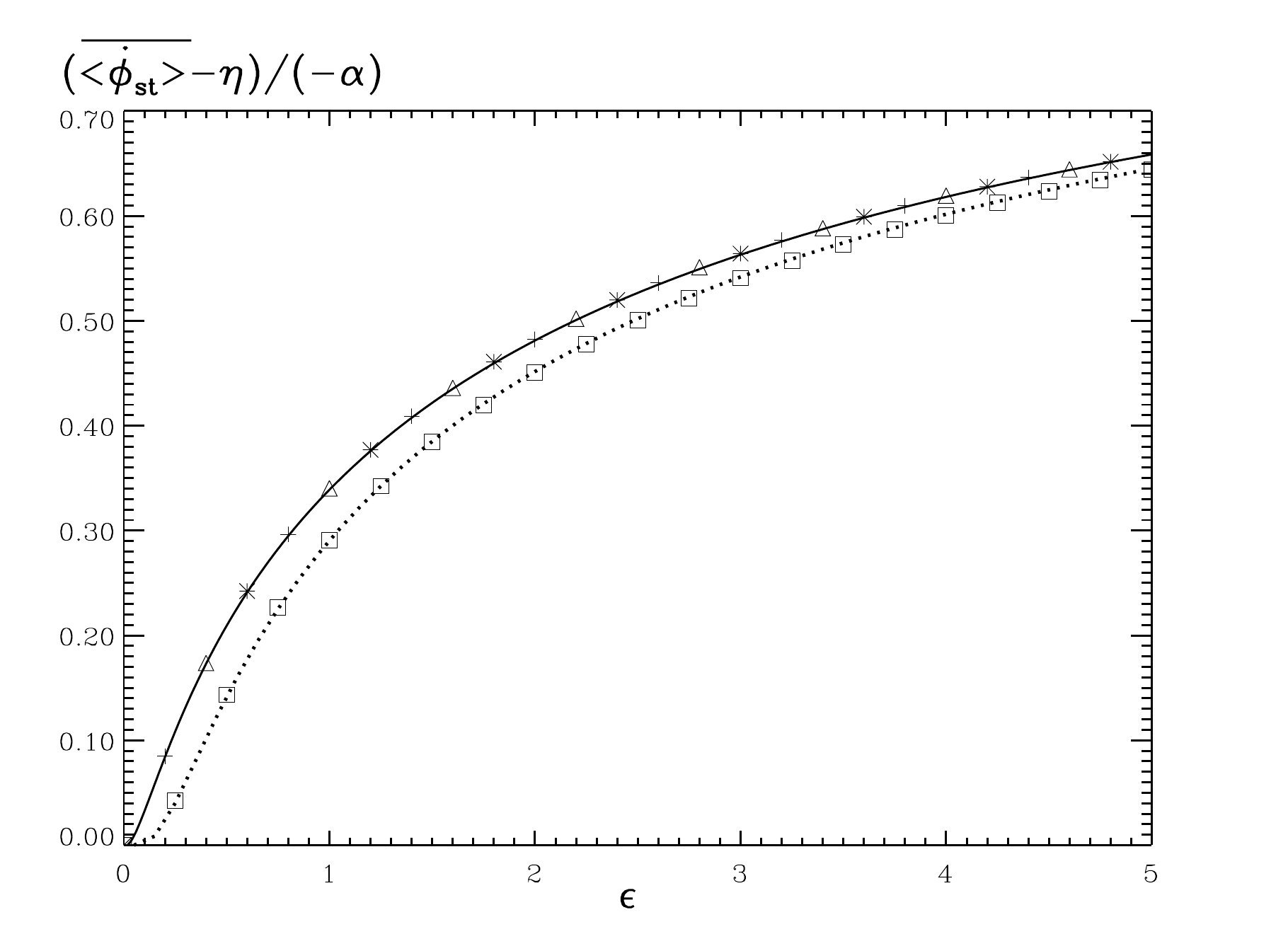}} \caption{Plot of the normalised averaged frequency $[{\overline{\langle \dot{\phi}\rangle_\textrm{st}}-\eta}]/{\alpha}$
 as a function of the noise intensity $\epsilon$. This normalised frequency neither depends on $\eta$ nor on $\alpha$.
 The lines correspond to the theoretical expression obtained using (\ref{dphasemean}). The symbols are the results of averaging over at time $10^5 T$ the data for $\dot\phi(t)$ coming from the numerical integration of Eqs.(\ref{cAm:x1},\ref{cAm:x2}). Same parameters and symbol meaning that in Fig.\ref{figureintensity}.}\label{figurefreq}
\end{figure}

We note that the effect of $\eta$ is to add a constant value $\eta$ to $\overline{\langle \dot{\phi} \rangle_\textrm{st}}$. Additionally, $\overline{\langle \dot{\phi} \rangle_\textrm{st}}$ depends linearly on $\alpha$. For these reasons we define the normalised averaged frequency $[\overline{\langle \dot{\phi} \rangle_\textrm{st}} -\eta]/(-\alpha)$ that is constant for different values of $\alpha$ and $\eta$, see Fig.\ref{figurefreq}.

For $\rho=0$ we can use
the analytical Eqs.(\ref{Ist0},\ref{phist0}) with the explicit dependence on time. In this case,
we can obtain the limits for vanishing or extremely large noise. For $\epsilon \rightarrow 0, \overline{\langle I \rangle_\textrm{st}} \approx a_0-b$ and $\overline{\langle \dot{\phi} \rangle_\textrm{st}} \approx \eta$, hence $[\overline{\langle \dot{\phi} \rangle_\textrm{st}} -\eta]/(-\alpha)\approx 0$. For $\epsilon \rightarrow \infty, \overline{\langle I \rangle_\textrm{st}} \approx a_0 + 2 \epsilon$ and $\overline{\langle \dot{\phi} \rangle_\textrm{st}} \approx -\alpha+\eta$, then $[\overline{\langle \dot{\phi} \rangle_\textrm{st}} -\eta]/(-\alpha)\approx 1$. These tendencies are observed in Figs.\ref{figureintensity} and \ref{figurefreq}.

We want to compare these theoretical expressions with the mean values obtained numerically. By integrating Eqs.(\ref{cAm:x1},\ref{cAm:x2}), getting the corresponding intensity $I(t)$ and phase $\phi(t)$ after the transient dynamics, we can obtain the numerical mean values of $\overline{\langle I \rangle_\textrm{st}}$ and $\overline{\langle \dot{\phi} \rangle_\textrm{st}}$ near the minima of the potential. Whenever the two evaluation procedures (theory and numerics) give the same
result it will imply that $P_\textrm{st}(I,\phi,t)$,
the above indicated probability distribution function, is the adequate one. In Figs.\ref{figureintensity} and \ref{figurefreq} we compare the numerical results (symbols) with the theoretical ones (lines) obtaining
a very good agreement. Consequently the potential (\ref{vi}) used in the model without modulation extents its validity into the modulation case.

The effect of $\rho$ is observed in Figs.\ref{figureintensity} and \ref{figurefreq}, dotted lines for $\rho\ne0$. Both $\overline{\langle I \rangle_\textrm{st}}$ and $\overline{\langle \dot{\phi} \rangle_\textrm{st}}$ depend on $\rho$. For small values of $\epsilon$ the effect of $\rho$ is to increase the mean value of the intensity and decrease the absolute value of the frequency. For $\rho \ne 0$ the potential is tilted in a preferred direction, hence its minimum has a larger value of the intensity and the absolute value of the frequency is lower because it is more difficult to escape from the minimum. However, for large values of the noise term $\epsilon$, the trajectories are far from the minimum and the dynamics is similar to the one for $\rho=0$. In fact, for $\rho \ne 0$ and $\epsilon \rightarrow \infty$, the mean values have the same tendency than in the $\rho=0$ case, this is $\overline{\langle I \rangle_\textrm{st}} \approx a_0 + 2 \epsilon$ and $[\overline{\langle \dot{\phi} \rangle_\textrm{st}} -\eta]/(-\alpha)\approx 1$.

\section{Stochastic Resonance: the Amplification Factor}

It is the aim of this section to study if a certain value of the noise term $\epsilon$ improves the response of some of the variables for the modulated system. More specifically, we are interested in maximising the amplification factor of the intensity or the phase flux as a function of the noise term.

The amplification factor of $\langle I \rangle_\textrm{st}(t)$ is defined as
\begin{equation}\label{afi}
A_F (I)\equiv 4 |M_1|^2/{a_1}^2,
\end{equation}
where $M_1$ indicates the
first coefficient of the Fourier expansion~\cite{juhae}
\begin{equation}
\langle I \rangle_\textrm{st}(t) = \sum_n M_n \exp{(i n \omega t)},
\end{equation}
and can be obtained as
\begin{equation}\label{m1i}
M_1 = \frac{\omega}{2 \pi} \int_0^{\frac{2 \pi}{\omega}} \langle I \rangle_\textrm{st}(t) \exp{(-i \omega t)} dt.
\end{equation}
We evaluate this amplification factor by using $\langle I \rangle_\textrm{st}(t)$ in two ways: (i) by direct numerical integration of Eqs.(\ref{cAm:x1},\ref{cAm:x2}), and using as $\langle I \rangle_\textrm{st}(t)$ the mean value of $I(t)$ obtained in a single trajectory for different values of time, and (ii) by using the theoretical potential and computing $\langle I \rangle_\textrm{st}(t)=\int{I P_\textrm{st}(I,\phi,t)dI}$ as explained in the previous section. We define the amplification factor $A_F (\dot{\phi})$ of the frequency ($\dot{\phi}$) using an equivalent procedure.

Again, we note that the parameter $\omega$ does not have any influence in the amplification factors as the change of variables $t'=\omega t$ eliminates $\omega$ from all relevant integrals.

\begin{figure}[h]
\centerline{
\includegraphics[scale=0.5]{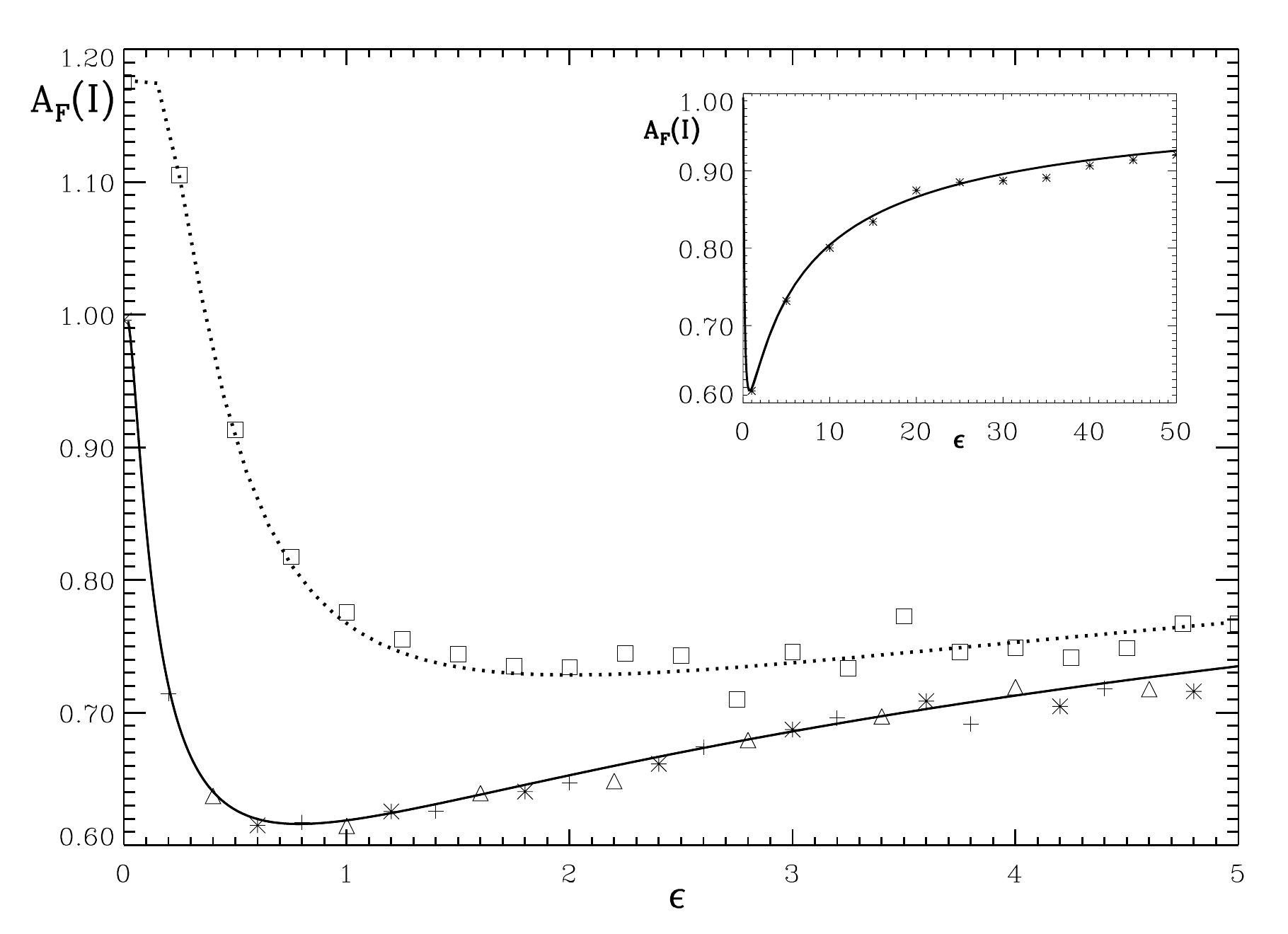}} \caption{Plot  of the amplification factor of the intensity $A_F(I)$ as a function of the noise intensity $\epsilon$, obtained from Eqs.(\ref{afi},\ref{m1i}), the data for $\langle I\rangle_\text{st}(t)$ coming either from the analytical expression Eq.(\ref{Ist}) with a time-dependent potential (lines), or from the numerical integration of Eqs.(\ref{cAm:x1},\ref{cAm:x2}) (symbols). In the main plot, the symbols are the results of averaging over a time $10^5 T$ while in the inset, $\alpha=2$, $\eta=0$ and $\rho=0$, the averages are performed during a time period of $10^7 T$. Same parameters and symbol meanings than in Fig.\ref{figureintensity}. }\label{figurefactintensity}
\end{figure}

\begin{figure}[h]
\centerline{
\includegraphics[scale=0.5]{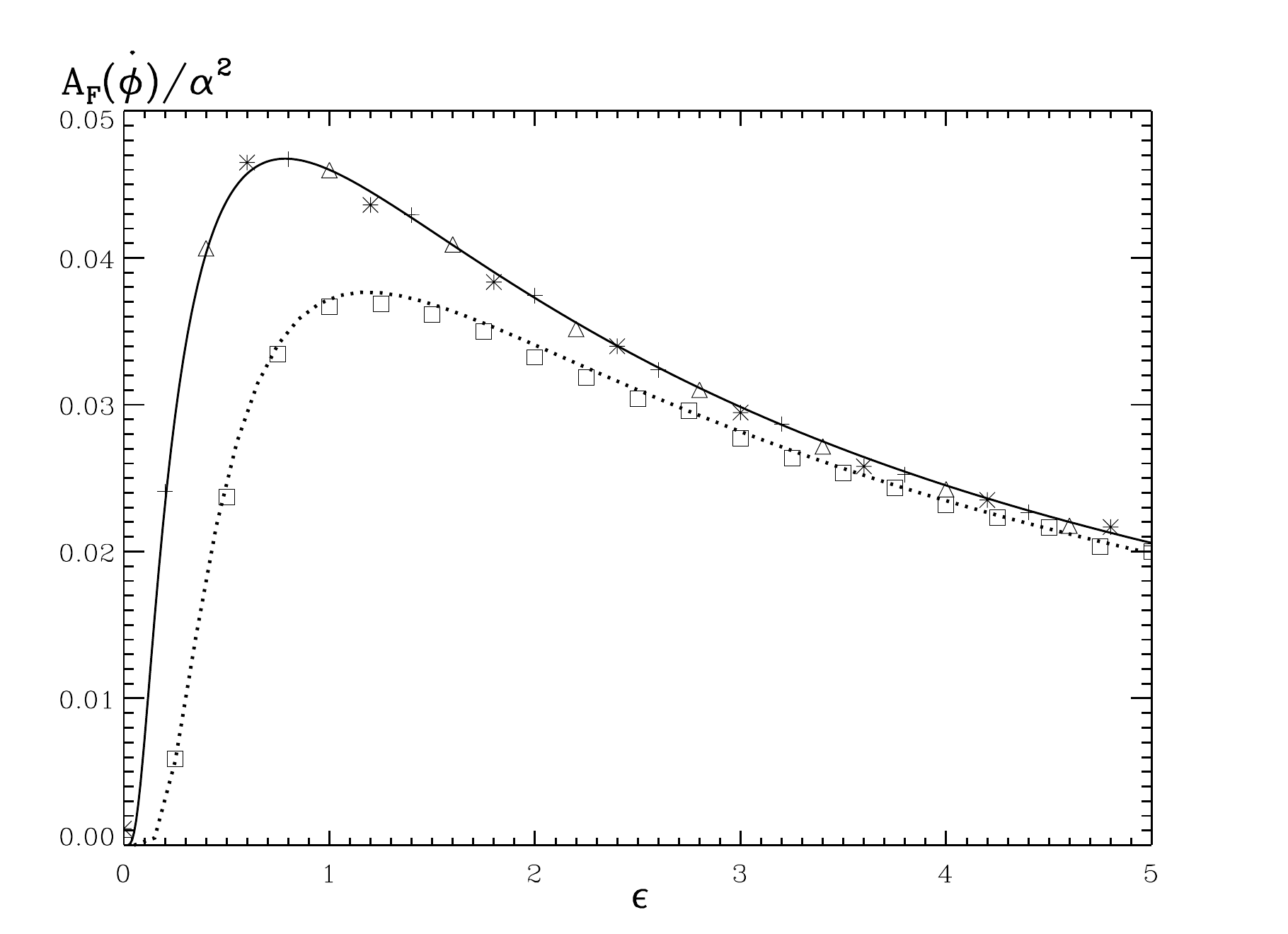}} \caption{Plot of the normalised amplification factor of the frequency $A_F(\dot{\phi}) /\alpha^2$ as a function of the noise intensity $\epsilon$. This particular normalisation does neither depends on $\alpha$, nor on $\eta$. The lines and symbols are obtained by a similar procedure as explained in the caption of Fig.\ref{figurefactintensity} using the frequency $\dot\phi$ instead of the intensity $I$. Same parameters and symbol meanings than in Fig.\ref{figureintensity}.}\label{figurefactfreq}
\end{figure}

In Figs. \ref{figurefactintensity} and \ref{figurefactfreq} we plot the above defined amplification factors as a function of $\epsilon$. We obtain a good agreement between the results obtained with the use of the potential function or by a direct numerical integration of the dynamical equations, indicating the consistency of the theoretical frame (compare lines and symbols of both figures). In the numerical simulations, and specially for large $\epsilon$, it is important to evaluate the mean values over longer times in order to reduce the statistical errors and obtain a good agreement between the numerical results and the theoretical ones, as it is observed in the inset of Fig.\ref{figurefactintensity}. For $\rho=0$ we can use the analytical expressions (\ref{Ist0}) and (\ref{phist0}) with the explicit dependence on time. In this case, we can obtain the limit for both vanishing and very large noise. For $\epsilon \rightarrow 0$ and for $\epsilon \rightarrow \infty$, both $A_F(I) \rightarrow 1$ and $A_F(\dot{\phi}) \rightarrow 0$.

Let us explain now the main effect of the parameters $\alpha$, $\eta$ and $\rho$ on the amplifications factors, as evidenced in Figs. \ref{figurefactintensity} and \ref{figurefactfreq}.

The parameter $\alpha$ has no effect when evaluating theoretically  the amplification factor $A_F(I)$ because it appears in
the antisymmetric matrix but it does not explicitly appear in the potential. The amplification factor $A_F(\dot{\phi})$ does have
an explicit dependence on $\alpha$, and we can define a normalised amplification factor $A_F(\dot{\phi})/\alpha^2$ that does not depend on $\alpha$.

From our theoretical expressions, we obtain that the injected signal frequency $\eta$ does not have any influence neither in the values of $A_F(I)$ nor on $A_F(\dot{\phi})$, as it corresponds to a residual term in the dynamics in terms of the potential.

The amplitude $\rho$ of the injected field increases the amplification factor of the intensity (compare the dotted and solid lines of Fig.\ref{figurefactintensity}) and decreases the amplification factor of the frequency (compare the dotted and solid lines of Fig.\ref{figurefactfreq}). For $\rho>0$ the potential, tilted to a preferred direction, makes the system
to evolve only around one well. For large values of the noise intensity $\epsilon$, the dynamical variables evolve in the
upper part of the potential, which is the same than for $\rho =0$. Consequently, the amplification
factor of the frequency coincides with the one for $\rho=0$, as said before, for $\epsilon \rightarrow \infty$, $A_F(I) \rightarrow 1$ and $A_F(\dot{\phi}) \rightarrow 0$.

We now discuss the contribution of the parameters $a_0$, $a_1$ related with the modulation terms (for the sake of brevity we do not provide specific figures sustaining the next statements).

If $a_0$ increases (for fixed values of $b$ and $a_1$), both the mean value of the intensity and the amplification factor $A_F(I)$ increase, while the absolute value of the frequency and the corresponding amplification factor $A_F(\dot{\phi})$ decrease. In terms of the potential, the larger $a_0$, the wider the diameter of the circle of the minima of the "Mexican hat", hence the higher the intensity and the smaller the frequency. A similar effect would be obtained when $a_0$ remains constant and $b$ is decreased, as the potential's minimum is related to $a_0-b$.

The amplification factor does not change significantly when $a_1$ is modified (keeping $a_0$ and $b$ constant). However, we observe that when $a_1$ increases, the mean value of the intensity slightly increases but the amplification factor $A_F(I)$ decreases, and the absolute value of the frequency and the amplification factor $A_F(\dot{\phi})$ increase.

An important result is that the amplification factor $A_F(I)$ as a function of $\epsilon$ has a minimum, while the amplification factor $A_F(\dot{\phi})$ presents a maximum. Moreover, for $\rho=0$ the minimum of $A_F(I)$ and the maximum for $A_F(\dot{\phi})$ occur at the same value of the noise parameter $\epsilon$, while for $\rho>0$ the coincidence between the location of the two extrema is not so accurate. These extreme values correspond to the value of $\epsilon$ for which the system goes out of the influence of the minima of the potential. This noise intensity can be obtained numerically from equation (\ref{vi}) identifying the value of $\overline{\langle I \rangle_\textrm{st}}$ which has the same value of the potential that the one of the relative maximum of the potential $I_{max}=0$. For $\rho=0$, $V(\overline{\langle I \rangle_\textrm{st}})=V(I=0)$, equivalent to the equation $\overline{\langle I \rangle_\textrm{st}} - a \ln(1+\overline{\langle I \rangle_\textrm{st}}/b)=0$. For the values we consider in Figs.\ref{figurefactintensity} and \ref{figurefactfreq}, we obtain a value of $\overline{\langle I \rangle_\textrm{st}} \approx 2.5$, which corresponds from Fig.\ref{figureintensity} to a noise term $\epsilon \approx 0.75$. This value agrees to the observed one for which there is minimum of the amplification factor for the intensity and a maximum of the amplification factor of the frequency.
For $\rho \ne 0$, the condition $V(\overline{\langle I \rangle_\textrm{st}})= V(I=0)$ leads to an equation that depends on $\phi$ and we take the value of $\phi$ that maximises the intensity,  namely $\overline{\langle I \rangle_\textrm{st}} -a \ln(1+\overline{\langle I \rangle_\textrm{st}}/b)-\dfrac{2\rho\sqrt{\overline{\langle I \rangle_\textrm{st}}}}{\sqrt{1+\alpha^2}}=0$.
For the values we consider in Figs.\ref{figurefactintensity} and \ref{figurefactfreq}, we get $\overline{\langle I \rangle_\textrm{st}} \approx 6.1$, which corresponds from Fig.\ref{figureintensity} to a noise term $\epsilon \approx 2.1$, which agrees reasonably with the value for which the amplification factor of the intensity displays a minimum, see Fig.\ref{figurefactintensity}. For larger values of the $\rho$ parameter, i.e. $\rho=5$, the minimum of the amplification factor of the intensity disappears whereas the maximum of the amplification factor of the frequency is kept.

\section{Conclusions}

In this paper we have exploited the description of a class A laser in terms of a Lyapunov potential for the deterministic dynamics
in order to discuss the effect of a modulation term in the gain parameter. The potential description was
obtained in a previous work~\cite{laser0,laser1} and is strictly valid in the $\rho\eta =0$ parameter region, but can be extended approximately to other cases. For the unmodulated case, the use of the potential function allows one to derive analytically expressions for the dependence of the mean intensity and phase flux on the system parameters, including the noise intensity. For the modulated case, we are also able to evaluate the amplification factors of the intensity and phase flux by a numerical integration of the equations. Different values for the parameters describing
the system were considered and the results compared with the theoretical
ones. Stochastic resonance is obtained as an indication that it is possible to have a maximum response for the frequency of the system whenever the noise term is chosen properly. The agreement of both results validates the use of the Lyapunov potential for the case of a modulated laser.

\section{Acknowledgements}
We acknowledge financial support from Agencia Estatal de Investigaci\'on (AEI, Spain) and Fondo Europeo de Desarrollo Regional under Project ESoTECoS Grant No. FIS2015-63628-C2-2-R (AEI/FEDER,UE). This work was started when HSW was at IFCA (UC-CSIC).

\newpage

\begin{thebibliography}{90}

\bibitem{sr-01} L. Gammaitoni, P. H\"{a}nggi, P. Jung, F. Marchesoni, Rev. Mod. Phys. \textbf{70}, 223 (1998)

\bibitem{sr-02} T. Wellens, V. Shatokin, A. Buchleitner, Rep. Prog. Phys. \textbf{67}, 45 (2004)

\bibitem{abott} M.McDonell, N.G.Stocks, Ch.E.M.Pierce, D. Abbott;
From Suprathreshold Stochastic Resonance to Stochastic Signal Quantization (Cambridge University Press, 2008)

\bibitem{sr-04} J.K. Douglas et al., Nature \textbf{365}, 337 (1993); J.J. Collins et al., Nature \textbf{376}, 236 (1995);
S.M. Bezrukov, I. Vodyanoy, Nature \textbf{378}, 362 (1995)

\bibitem{sr-05} A. Guderian, G. Dechert, K. Zeyer, F. Schneider, J. Phys. Chem. \textbf{100}, 4437 (1996); A. F\"{o}rster,
M. Merget, F. Schneider, J. Phys. Chem. \textbf{100}, 4442 (1996); W. Hohmann, J. M¨uller, F.W.
Schneider; J. Phys. Chem. \textbf{100}, 5388 (1996)

\bibitem{sr-06} J.F. Lindner et al., Phys. Rev. E \textbf{53}, 2081 (1996)

\bibitem{sr-07} A.R. Bulsara, G. Schmera, Phys. Rev. E \textbf{47}, 3734 (1993); P. Jung, U. Behn, E. Pantazelou,
F. Moss, Phys. Rev. A \textbf{46}, R1709 (1992); P. Jung, G. Mayer-Kress, Phys. Rev. Lett. \textbf{74}, 2130
(1995); J.F. Lindner et al., Phys. Rev. Lett. \textbf{75}, 3 (1995); F. Marchesoni, L. Gammaitoni, A.R.
Bulsara, Phys. Rev. Lett. \textbf{76}, 2609 (1996)

\bibitem{sr-08} H.S. Wio, Phys. Rev. E \textbf{54}, R3075 (1996); H.S. Wio, F. Castelpoggi, Unsolved Problems of
Noise, Proc. Conf. UPoN'96, edited by C.R. Doering, L.B. Kiss, M. Schlesinger (World Scientific,
Singapore, 1997), p. 229; F. Castelpoggi, H.S. Wio, Europhys. Lett. \textbf{38}, 91 (1997)

\bibitem{sr-09} F. Castelpoggi, H.S. Wio, Phys. Rev. E \textbf{57}, 5112 (1998);
H.S. Wio et al., Physica A \textbf{257}, 275 (1998); M. Kuperman, H.S. Wio, G. Iz\'us, R. Deza, Phys. Rev.
E \textbf{57}, 5122 (1998)

\bibitem{sr-10} H.S. Wio et al., Physica A \textbf{257}, 275 (1998); M. Kuperman, H.S. Wio, G. Iz\'us,
R. Deza, Phys. Rev. E \textbf{57}, 5122 (1998)

\bibitem{sr-11} S. Bouzat, H.S. Wio, Phys. Rev. E \textbf{59}, 5142 (1999)

\bibitem{sr-12} B. von Haeften, R. Deza, H.S. Wio, Phys. Rev. Lett. \textbf{84}, 404 (2000)

\bibitem{sr-13} C.J. Tessone, H.S. Wio, P. H\"{a}nggi, Phys. Rev. E \textbf{62}, 4623 (2000)

\bibitem{sr-14} M.A. Fuentes, R. Toral, H.S. Wio, Physica A \textbf{295}, 114 (2001)

\bibitem{Graham} Graham R., in \textit{Instabilities and Nonequilibrium Structures},
Eds. E.Tirapegui and D.Villaroel (D.Reidel, Dordrecht, 1987).

\bibitem{sr-15} H.S. Wio, S. Bouzat, B. von Haeften, in Proc. 21st IUPAP Int. Conf.
on Statistical Physics,
STATPHYS21, edited by A. Robledo, M. Barbosa, Physica A \textbf{306C}, 140 (2002)

\bibitem{laser0} C. Mayol, R. Toral, and C. R. Mirasso, Phys. Rev. A \textbf{59}, 4690 (1999)

\bibitem{laser1} C. Mayol, R. Toral, C. R. Mirasso and M. A. Natiello,
Phys. Rev. A \textbf{66}, 013808 (2002)

\bibitem{las1} J. R. Tredicce, G. L. Lippi, and G. P. Puccioni, J. Opt. Soc.
Am. B \textbf{2}, 173 (1985).

\bibitem{las2} H. G. Solari and G. L. Oppo, Opt. Commun. \textbf{111}, 173 (1994).

\bibitem{las3} S. Wieczork, B. Krauskopf, and D. Lenstra, Opt. Commun.
\textbf{172}, 279 (2000).

\bibitem{las4} T. Simpson, J. M. Liu, K. F. Huang, and K. Tai, Quantum
Semiclassic. Opt. 9, \textbf{765} (1997).

\bibitem{las5} H. Haken, \textit{Laser Light Dynamics}, Light Vol. 2
(North-Holland, Amsterdam, 1985).

\bibitem{haken} H. Haken, \textit{Laser Theory} (Springer-Verlag, New York, 1984).

\bibitem{vandergraaf} W. A. van der Graaf, Ph.D. thesis, Vrije Universiteit, Amsterdam, 1997.

\bibitem{Agrawal} G. P. Agrawal and N. K. Dutta, Long-Wavelength Semiconductor Lasers (Van Nostrand, Reinhold, New York, 1986).

\bibitem{hsw} H.S. Wio, in \textit{4th Granada Seminar in Computational Physics},
P. Garrido, J. Marro, Eds. (Springer-Verlag, Berlin, 1997), p. 135

\bibitem{wdl} H.S. Wio, R.R. Deza and J.M. L\'opez, \textit{Introduction to Stochastic
Processes and Nonequilibrium Statistical Physics}, Revised Edition (World Scientific,
Singapore, 2013).

\bibitem{Gardi} C. W. Gardiner, \textit{Handbook of Stochastic Methods},
Eds. 4th (Springer-Verlag, Berlin, 2009)

\bibitem{tabla} M. Abramowitz and I. A. Stegun, \textit{Handbook of Mathematical Functions:
with Formulas, Graphs, and Mathematical Tables}, (Dover, New York, 1965).

\bibitem{juhae} P. Jung and P. H\"{a}nggi, Phys. Rev. A \textbf{44}, 8032 (1991)

\end{thebibliography}
\end{document}